\pdfoutput=1
\documentclass[12pt,notitlepage]{article}
\usepackage{times}

\usepackage{times}
\usepackage{mathptmx}

\newcommand{\xx}[1]{\section {#1}}
\newcommand{\xxx}[1]{\subsection {#1}}

\newcommand{\vardef}[3]{\item $ {#1} $ is the {#2}. }

\providecommand{\spacing}[1]{\renewcommand{\baselinestretch}{#1}\small\normalsize}
\providecommand{\doublespacing}{\spacing{1.5}}
\providecommand{\normalspacing}{\spacing{1.0}}

\newcommand{\eql}[2]{\begin{equation} \label{eq:#1} #2 \end{equation}}
\newcommand{\eqr}[1]{Eq.~(\ref{eq:#1})}

\newcommand{\secl}[1]{\label{sec:#1}}
\newcommand{\secr}[1]{\S\,\ref{sec:#1}}

\newcommand{\tcaption}[2]{\caption{\label{tab:#1} {#2}}}
\newcommand{\tabr}[1]{Table~\protect{\ref{tab:#1}}}

\usepackage[normalem]{ulem}

\newcommand{\dpdp}[2]{\mbox{$ \frac{\partial #1}{\partial #2} $}}
\newcommand{\dvar}[1]{{#1}d-$\!$VAR}
\newcommand{\eg}{{\em e.g.}}

\newcommand{\ie}{{\em i.e.}}

\newcommand{\wrt}{with respect to}

\renewcommand{\vardef}[3]{\item $ {#1} $ is the {#2} [{#3}]. }

\usepackage{ametsoc}

\newcommand{\ssfootnote}[1]{\footnote{\normalspacing\footnotesize {#1}}}

\begin{document}

\title{A retrieval strategy for interactive ensemble data assimilation}

\author{Ross N. Hoffman\footnotemark[1]}

\footnotetext[1]{\em Contact information:
\upshape
Dr. Ross N. Hoffman,
Atmospheric and Environmental Research, Inc.,
131 Hartwell Avenue,
Lexington, MA 02421-3126
Email: ross.n.hoffman@aer.com.}

\maketitle


\xx {Introduction}

\stepcounter{footnote}

Is it better to use radiances or retrievals in data assimilation?
I started out my career believing in radiance assimilation
\citep{Hof83,HofN89}, but now, with radiance assimilation having
become mainstream, I'm having second thoughts.
Here I outline an approach following \citet{Rod00} that might have
some advantages that uses retrievals along with the so-called Jacobian
or sensitivity matrix, actual averaging kernel (AK), actual prior, and
estimated retrieval noise covariance.
(See \secr{AK}.)
Others have considered using retrievals along with other information
provided by the retrieval process \citep{Joid98}.
Eyre and coworkers used \dvar1 retrievals within \dvar3 and
\dvar4 assimilation systems \citep{EyrKM+93,SzyCE04,PavEE08}.

Using retrievals gives us several advantages \citep[Section 8.3]{Rod00}.
First it reduces complexity in the $H$-operator, reduces data volume,
allows sequential processing within the retrieval algorithm, allows
arbitrary cloud clearing methods, and makes the assimilation system more
modular.
(Sequential processing means, for example, first use one band for
temperature retrievals, then use some other band for humidity
retrievals, etc.)\ssfootnote{Why do retrieval people like to do this?
I think because of uncertainty in the spectroscopy and in estimating
errors, and because the optimal coupled problem is so sensitive.}
(More in \secr{pro-con}.)
Using EOFs gives us additional advantages, including reduced data
volume. (See \secr{EOFs}.)
There is also the potential to reduce vertical interpolation errors
that arise due to the different vertical grids used in the forward
problem and the meteorological model. (See \secr{EOF-vinterp}.)
Note that if the retrieval method uses an EOF representation, then
formally the error covariance matrix in physical space is not full
rank.
(But see below in the discussion of ``EOF covariances'' where we
describe how we create the physical space error covariance matrix at
AER.)
Finally the AK provides the opportunity to remove the influence of the
prior, and allows for interactive retrievals (where, for example, the
prior is the ensemble mean forecast). (See \secr{interactive}.)

The motivation for this approach is for instruments with thousands of
channels.
This includes AIRS, IASI and the Aura TES.
In these cases, instead of using thousands of channels or having to
decide which are the optimal channels to use in the DAS, these issues
would be pushed back to the retrieval scheme.
But this approach can be applied more generally, and has the advantage
of explicitly treating correlations of errors for a single
retrieved profile, with only modest impacts on the DAS.

There are three successive developments in what follows.
First, in \secr{AK}, we show how to use the AK and the retrieval noise
covariance to transform the retrieved quantities into observations
that are unbiased and have uncorrelated errors, and to eliminate
both the smoothing inherent in the retrieval process and the effect of
the prior.
Second, in \secr{EOFs}, we show how to transform this result into EOF
space, when a truncated EOF series has been used in the retrieval
process.
This provides a degree of data compression and eliminates those
transformed variables that have very small information content.
In both approaches a vertical interpolation from the dynamical model
coordinate to the radiative transfer coordinate is required.
In the third development, in \secr{EOF-vinterp}, we propose using the
EOFs to essentially reduce the error of the vertical interpolation.

\xx {Radiance assimilation :: pros and cons\secl{pro-con}}

The advantages of direct radiance assimilation are that we use the
data in its basic state and there are weak or no error correlations.
The disadvantages are that there may be huge numbers of channels, and
that what I've termed geophysical biases can effectively induce
observation error correlations.
For example, the meteorological model might have a very simple
representation of surface emissivity, while the true surface emissivity
is governed by geologic parameters that have long length scales.
This difference must be considered an observational error in data
assimilation systems (DAS).\ssfootnote{Like ``fish'' [one fish, two
fish \dots] ``DAS'' is both singular and plural.}
A retrieval scheme might employ a slack variable (\eg, cloud
fraction) or use information from some channels to model the
emissivity in other channels.
Further, spectroscopic inconsistencies may introduce correlations in
the forward problem, which enter the problem in the same way as
observational error correlations.
In the end, I expect that whatever can be done in a retrieval scheme can
probably be done in DAS if we are clever enough, but at a cost in
complexity and maintenance.
I believe it is best to avoid the use of radiances in the DAS and
instead use retrievals with a strategy along the lines outlined here
whenever an intricate retrieval schemes is favored by the remote
sensing community to extract the maximum information content from the
observations.

\xx {Notation}

As we will quote so many results from \citet{Rod00}, we will adopt his
notation, defining terms as we go.
However, this means that $x$ discussed here, the vector of retrieved
quantities\ssfootnote{Typically a temperature profile.}, become the
observations, usually denoted $y$ in an EnKF or \dvar4 DAS context,
while at the same time $y$ in \citet{Rod00} denotes the channel
radiances.
Moreover in what follows we propose that we use a transformation of
$x$ that involves the AK ($A$) as an observation in the DAS, so to
bridge the notation gap, we will denote this transformation as $y_A$.

\xx {Averaging kernel method\secl{AK}}

In theory \citep[\eg,][]{Rod00}, a significant fraction of retrieval error
results from the fact that the retrievals can not observe the detailed
vertical structure of the atmosphere and prior information is used to
fill some of the gaps, but the result is smoothed compared to reality.
When used as a profile, this corresponds to correlated errors.
Also, using the best possible prior (the forecast!) is considered a
no-no by many.
\citet[Section 8.3.4]{Rod00} suggests a one-sentence way around this.
He leaves a few details for the student to fill in, but most of what
we need is scattered throughout his excellent book.
To begin, \citet[Eq.\ 3.12]{Rod00} writes the retrieval as 
\eql{xhat}{\hat{x} = A x + (I-A) x_a + G_y \epsilon_y.}
Here $\hat{x}$ is the retrieval, $x$ is the true state vector, and
$x_a$ is the prior.
The matrix $A$ is the averaging kernel.
The total measurement error relative to the forward problem,
$\epsilon_y$ \citep[Eq.\ 3.11]{Rod00}, includes measurement error and
forward model error and is usually characterized as
being unbiased and having a specified covariance matrix,
$S_{\epsilon}$.\ssfootnote{In practice $S_{\epsilon}$ is often taken
to be diagonal, but that is not material in this analysis. We return
to a discussion of $S_{\epsilon}$ in \secr{forwardErrors}.}
In \eqr{xhat} the matrix $G_y$ is the sensitivity of the retrieval to
the radiances and is given by \citet[Eqs. 3.25, 3.27]{Rod00} as
\eql{Gy1}{G_y = \dpdp{\hat{x}}{y} = \hat{S} K^T S_{\epsilon}^{-1}.}
Here $K=\dpdp{F}{x}$ is the Jacobian of the forward problem, $F$,
evaluated at the solution $\hat{x}$ \citep[Eq.\ 2.20]{Rod00}, and 
\eql{Shat}{\hat{S}^{-1} = K^T S_{\epsilon}^{-1} K + S_a^{-1}}
defines the accuracy (inverse of the error covariance) of the
retrieval in terms of the accuracy of $G_y y$ and the accuracy of $x_a$
\citep[Eqs. 2.27, 4.13]{Rod00}.
An alternative computational
formula for $G_y$, given by \citet[Eqs. 2.45]{Rod00} is
\eql{Gy}{G_y = S_a K^T S_q^{-1}}
with
\eql{Sq}{S_q = K S_a K^T + S_{\epsilon}}
in which $S_q$ should be well conditioned due to the presence of
$S_{\epsilon}$.\ssfootnote{\citet[Section 4.1]{Rod00} demonstrates the algebraic
equivalence of these two forms.
In brief, find that $ \hat{S}^{-1} G_y S_q = K^T (I +
S_{\epsilon}^{-1} K S_a K^T)$ for both forms of $G_y$.}
With this notation we may write
\eql{AK}{A = G_y K = \hat{S} (K^T S_{\epsilon}^{-1} K) = I-\hat{S} S_a^{-1}}
following \citet[Eqs. 2.78, 2.79]{Rod00}.\ssfootnote{To see the
equivalence of the second two forms, note that
$I=\hat{S}\hat{S}^{-1}=\hat{S}(K^T S_{\epsilon}^{-1} K + S_a^{-1})$
according to \eqr{Shat}, and subtract the second form to find that
$I-A=\hat{S} S_a^{-1}$.}

Within a generalized DAS, where any observation can be used if it can
be calculated, this formulation of the retrieval can be used to
eliminate the impact of the prior.
For this purpose we posit a pre-processing step that removes the prior
and effectively removes smoothing as an error source \citep[Section
8.3.4]{Rod00} by defining the new observation as
\eql{yhatA1}{\hat{y}_A = \hat{x} - (I-A) x_a,}
and the new observation operator by
\eql{yA1}{y_A = A x.}
The retrievals are of course still smooth, but we will now be comparing
observed and simulated quantities with the same degree of smoothing.
Now the observation increments
\eql{obs-inc1}{\hat{y}_A - y_A = G_y \epsilon_y,}
are unbiased if the $\epsilon_y$ are unbiased, and have covariance
\eql{Sm}{S_m = G_y S_{\epsilon} G_y^T.}
\citet[Eq.\ 3.19]{Rod00} calls this the retrieval noise covariance.
An alternative form for \eqr{Sm} is
\eql{SmA}{S_m = A \hat{S},}
which follows from substituting \eqr{Gy1} into \eqr{Sm}, and then
making use of the second form of \eqr{AK}.\ssfootnote{Although $A$ is
not symmetric, $A \hat{S}$ is symmetric since $S_m$ is symmetric.
Explicitly $(A \hat{S})^T= \hat{S}^T A^T =
\hat{S} (K^T S_{\epsilon}^{-1} K)^T \hat{S} = \hat{S} (K^T
S_{\epsilon}^{-1} K) \hat{S} = A \hat{S}.$}

The end result is the same situation that is always obtained in
maximum \emph{a~posteriori} (MAP) retrieval schemes---you compare an
observation ($\hat{x} - (I-A) x_a$) to a simulated value ($Ax$) such
that the mean difference has zero expectation and you normalize the
difference by the square root of its inverse covariance
($S_m^{-\frac{1}{2}}$).

In this approach, the $H$-operator for this observation is simply $A
x$, but remember that $x$ is defined in the space of the forward
radiative transfer model.
Therefore the observation operator will in general require time
interpolation followed by interpolation in latitude and longitude,
followed by vertical interpolation to the vertical grid of the forward
problem and finally the computation of $y_A$ according to \eqr{yA1}.
Then, each element of the observation vector, $\hat{y}_A(i)$, will be
compared to the $i$th element of the simulated vector $y_A(i)$.
For convenience we might store the elements of the observation vector
individually, and compute the elements of the simulated vector
individually as the dot product of the $i$th row of $A$ and $x$.
Practically then, the individual elements of the observation vector, $\hat{y}_A(i)$,
might be stored along with the corresponding rows of $A$.
The rows of $A$ might also be used in determining vertical
localization.
Note that the preceeding arguments formally depend on the correctness of
\eqr{xhat}, including the implicit linearization assumption.

If we scale by $S_m^{-\frac{1}{2}}$ we effectively rotate to a space
where the observation errors are unbiased, uncorrelated, and have unit
variance.
Specifically we redefine the new observation as
\eql{yhatA}{\hat{y}_A = S_m^{-\frac{1}{2}} (\hat{x} - (I-A) x_a),}
and the new observation operator by
\eql{yA}{y_A = S_m^{-\frac{1}{2}} A x = A_R x,}
where we have defined the ``rotated'' AK,
\eql{AR}{A_R = S_m^{-\frac{1}{2}} A,}
for convenience.
Now the observation increments \eql{obs-inc}{\hat{y}_A - y_A =
S_m^{-\frac{1}{2}} G_y \epsilon_y,} are unbiased, and have covariance
$I$.\ssfootnote{Since $S_m$ is symmetric, we have $\langle (\hat{y}_A - y_A)
(\hat{y}_A - y_A)^T\rangle = S_m^{-\frac{1}{2}} G_y \langle\epsilon_y
\epsilon_y^T\rangle G_y^T S_m^{-\frac{1}{2}T} = S_m^{-\frac{1}{2}} S_m
S_m^{-\frac{1}{2}T} = I$.}

\xx {Forecast as prior\secl{interactive}}

After the pre-processing of \eqr{yhatA}, $x_a$ is no longer needed.
Since the observation increment does not depend on the prior, we can
use any prior.
One possibility is to use the ensemble background mean as $x_a$ and
the ensemble sample covariance as $S_a$.
We would call this the interactive retrieval approach.
This complicates the data flow since now the background ensemble
interpolated to the observation time, location, and vertical structure
must be provided to the retrieval scheme.
However, this is the best possible prior for the retrieval within the
context of data assimilation.
Further, if the system has settled down, then the ensemble background
mean should be a good estimate of the truth and the linearization
assumption of \eqr{xhat} should be a better approximation than otherwise.

\xx {Forward problem errors\secl{forwardErrors}}

The correct specification of the statistics of the combined
measurement and forward problem errors, $S_{\epsilon}$, is crucial to
the method.
Usually we think of $S_{\epsilon}$ as something specified by the
retrieval-performing organization.
However, there are two different approaches to determining
$S_{\epsilon}$, and the second one involves the data assimilation
system.
The first, what might be called a bottom-up approach, tries to estimate
all the sources of error separately and build up a composite estimate
of $S_{\epsilon}$.
The second, or top-down approach, is based on comparisons of radiances
and simulated radiances, most conveniently within the context of a
radiance data assimilation system.

In the bottom-up approach the following sources of error should be
considered:
\begin{itemize}
\item Measurement error including calibration error, misrepresentation
of the instrument response function, sources of stray radiation, geometry error,
geolocation error, polarization characterization, \ldots;
\item Spectroscopy errors (in line widths, strengths, continuum,
\ldots);
\item Neglect of NLTE effects;
\item Neglect or errors in how the magnetic field interacts with
radiation, including Zeeman line-splitting, Faraday rotation, \ldots;
\item Integration errors, i.e., truncation errors from the vertical grid;
\item EOF truncation errors;
\item Cloud clearing errors and beam-filling effects; and
\item Errors in physical properties that must be specified, but are not part of the control vector.
\end{itemize} The last item might include aerosol, cloud, and surface
properties as well as trace gas absorber amounts, such as CO$_2$
amounts, and the cosmic background in the microwave.
All of these effects could be estimated in simulation.
Estimating cloud clearing errors is likely the most difficult.

Among the top-down approaches, those derived from the method of
\citet{DesBC+05} seem most promising.
\citet{DesBC+05} combine differences among the observations, the
analysis, and the background, all in observation space to estimate the
analysis, background, and observation error covariances---again, all
in observation space.
To use this approach to estimate the forward problem error would
require the radiance observations and the background and analysis
evaluated as radiances.
For discussion here, let $x_b$ be the background for one of the
ensemble members
interpolated to the observation time and location, and to the vertical
grid of the forward problem.\ssfootnote{This interpolation is part of
our observation operator.}
Then we can estimate the corresponding background radiance $y_b$ as 
\eql{yb}{y_b = \hat{y} + K (x_b - \hat{x}),}
where $\hat{y} = F(\hat{x})$ is the radiance associated with the
retrieval.
The ensemble mean of $y_b$ would be the appropriate background in
observation space for the \citet{DesBC+05} method. 
The same approach applies to the analysis radiance, except to note
that the analysis is normally only available at the analysis time.
However, in the LETKF approach, the analysis weights can be applied at
each time (say each hour) to define the analysis at the same times as
the background.
Or the analysis weights could be applied directly to the $y_b$ to
define the analysis in observation space for the \citet{DesBC+05}
method.

Examples of studies that apply the so-called ``Desroziers diagnostic''
to radiance observation errors are
\citet{GarHB07,BorB10,BorCB10}.\ssfootnote{The later two references
also apply two other methods, the so-called Hollingsworth/Lonnberg
method \citep{HolL86} and a method based on subtracting a scaled
version of mapped assumed background errors from FG-departure
covariances.}
Within an enemble Kalman filter, \citet{LiKM09} demonstrate how it is
possible to continuously update both the observation and background
error statistics based on \citet{DesBC+05}.

For radiance data assimilation, the \citet{DesBC+05} approach includes
representativeness errors in the radiance errors.
If we don't want to include this error component, and \citet{Rod00}
would not, then we can project the observed radiances onto the
subspace sampled by the model.
To do this we would need to assume that the forward model is correct to
transfer model states to radiances, but then we would only use the
leading EOFs of a diverse sample of model states in radiance space to
filter the observations.

\xx {EOF properties\secl{EOF-properties}}

\textbf{Definition.}
Empirical orthogonal functions (EOFs) can be defined in terms of any
state vector.
For this description, suppose the state vector $x$ is the list of
temperatures at the levels used in the forward problem.
But $x$ could also include other profile parameters and surface
parameters with no loss of generality.
The EOFs are the eigenvectors of the $x$ sample-covariance matrix.
The EOFs are ordered by eigenvalue.
The variance explained by each EOF is given by its eigenvalue.
Thus, the first EOF is the anomaly ``pattern'' that occurs most frequently.

\textbf{Data compression.}
Since an EOF representation is often used in the retrieval to filter
small scales, the retrieval solution can be represented equivalently
(and with no loss of information) as a vector of EOF coefficient.
In the case of our AER retrievals we use a single set of EOFs globally
to avoid generating artificial edges in maps of the retrieved fields.
For Mars TES, 4-6 EOFs are sufficient.
Then the retrievals, error covariances, and AKs can be provided to the DAS in
terms of the leading EOFs.
If we reduce the size of the data vector from $n$ (say 30) to $j$ (say
5) we also reduce the size of the AK and error covariance matrices
from $O(n^2)$ to $O(j^2)$.

\textbf{EOF algebra.}
EOFs are used to relate anomalies, denoted $x'$, from an overall mean, 
$\bar{x}$.
That is,
\eql{xprime}{x' = x - \bar{x}.}
In any implementation we first subract $\bar{x}$ before converting to
EOF coefficients and then add it back when converting to physical
space.
The EOFs are truncated and ordered as columns in a matrix $E$.
In what follows it is only important that $E^TE = I$.
But note that if all EOFs are retained than $E^T$ and $E$ are inverses
so that $EE^T=I$ as well.

The projection of $x$ on the EOFs gives the EOF coefficients $\alpha$
according to \eql{projection}{E^Tx'=\alpha.}
Reconstruction and filtering (in the case where we have truncated the
EOF series) is obtained by \eql{filter}{E\alpha=x'_f.}
If we left multiple \eqr{filter} by $E^T$, then, since $E^TE = I$, we
find $E^Tx'_f=\alpha$.
Therefore $E^T(x'-x'_f)=0$, demonstrating that the part of $x'$ that
is filtered does not project onto the retained EOFs.

In summary, we made the following definitions:
\begin{itemize}
  \vardef{x'}{state vector}{K}
(Note that this is a deviation from an overall mean.)
  \vardef{E}{matrix of EOFs}{1}
  \vardef{\alpha}{vector of EOF coefficients}{K}
 \end{itemize}
We established these relationships:
\begin{itemize}
  \vardef{E^TE = I}{orthonormality relationship}{1}
  \vardef{E^Tx'=\alpha}{projection method}{K}
  \vardef{E\alpha=x'}{filter or reconstruction method}{K}
 \end{itemize}

\textbf{EOF covariances.}
The sample covariance matrix of $x$ is
\eql{ErrCov}{S=\left\langle(x-\langle x\rangle)(x-\langle
x\rangle)^T\right\rangle=E\left\langle(\alpha-\langle\alpha\rangle)-
(\alpha-\langle\alpha\rangle)^T\right\rangle E^T=E\tilde{S}E^T}
since $x-\langle x\rangle=(x-\bar{x})-(\langle x\rangle-\bar{x})=
E\alpha-E\langle\alpha\rangle$.
Here angle brackets denote an average over the sample.
If the sample is the posteriori distribution then $\langle x\rangle=\hat{x}$, if
the sample is the prior distribution then $\langle x\rangle=x_a$, and if the sample
is the climate distribution then $\langle x\rangle=\bar{x}$ and $\langle\alpha\rangle=0$.
If $E$ includes all eigenvectors, then $\tilde{S}=E^TSE$.
Generally we truncate the EOF series so $E$ is not full rank and
as a result, if the retrieval is done in terms of EOF coefficients,
then $\hat{S}$ will not be full rank either.
However, we can still apply $\tilde{S}=E^TSE$ in this case to obtain
the upper left corner of the full $\tilde{S}$ matrix as
desired.\ssfootnote{Partition $E$ into retained and truncated EOFs
$[E_r \; E_t]$ to see this.}
While \eqr{ErrCov} can be used to transform from EOF to physical space
covariance it is better to add in the ``noise'' present in the
truncated EOFs.
For example, in the AER retrieval we determine the error covariance matrix of the
EOF coefficients.\ssfootnote{This is a full symmetric matrix.}
Since the coefficients of the truncated EOFs are not estimated during
the retrieval we can add back the variability of these EOFs given by
the associated eigenvalues.
Computationally, take the diagonal matrix of eigenvalues, replace the
upper left block with the estimated error covariance of the EOF
coefficients from the retrieval scheme and apply \eqr{ErrCov} using
the full (not truncated) matrix $E$.

\xx {EOF analysis of the AK method\secl{EOFs}}

Note that \eqr{xhat} is usually written for the full physical space
quantities, but we can subtract $\bar{x} = A\bar{x} + (I-A)\bar{x}$
from \eqr{xhat} to get a version in terms of anomalies,
\eql{xhat-prime}{\hat{x}' = A x' + (I-A) x'_a + G_y \epsilon_y.}
We now project \eqr{xhat-prime} into EOF coefficient space.
To do this we filter each $x'$-vector.
For justification, note that the filtering step is normally done within
the retrieval algorithm initially and at each step of the iteration.
Since putting too much structure in the background for the retrieval
can degrade the results, it is also helpful to filter the prior.
Further, within the context of a retrieval in terms of EOF coefficients,
that part of the true $x'$ that does not project on the EOFs should be
considered part of the representativeness error.

To filter each $x'$-vector we simply replace each $x'$ by its filtered
version $EE^Tx'$.
Then, to project into the EOF space, we left multiply each equation by $E^T$.
In the case of \eqr{xhat-prime} we obtain
\eql{AK-ET}{(E^TE)(E^T\hat{x}') = (E^TAE)(E^Tx') +
[E^T(I-A)E](E^Tx'_a) + E^T G_y \epsilon_y,}
where we have collected terms in parentheses to indicate how to
apply the above relationships.
If we define \eql{tildeA}{\tilde{A} = E^TAE} then we have
\eql{AK-alpha}{\hat{\alpha} = \tilde{A}\alpha + (I-\tilde{A}) \alpha_a
+ \epsilon_{\alpha},}
where
\eql{epsilon-alpha}{\epsilon_{\alpha} = E^T G_y \epsilon_y,}
\eql{project-xhat}{\hat{\alpha} = E^T(\hat{x}-\bar{x}),} and
\eql{project-xa}{\alpha_a = E^T(x_a-\bar{x}).}
We may say that $\tilde{A}$ is the projection of $A$ into the EOF
coefficient space, or simply that $\tilde{A}$ is the EOF AK, and that
$\epsilon_{\alpha}$ is the projection of the retrieval error into the
EOF coefficient space, or simply that $\epsilon_{\alpha}$ is the EOF
retrieval error.

Now $\epsilon_{\alpha}$ has covariance given by
\eql{StildeM}{\tilde{S}_m = E^T G_y \langle \epsilon_y \epsilon_y^T
\rangle G_y^T E = E^T G_y
S_{\epsilon} G_y^T E = E^T S_m E .}
As we did before (in \secr{AK}), we scale by $\tilde{S}_m^{-\frac{1}{2}}$
to rotate to a space where the observation errors are unbiased,
uncorrelated, and have unit variance.
Specifically we redefine the new observation as
\eql{yhatA-EOF}{\hat{y}_A = \tilde{S}_m^{-\frac{1}{2}} (\hat{\alpha} - (I-\tilde{A}) \alpha_a),}
and the new observation operator by
\eql{yA-EOF}{y_A = \tilde{S}_m^{-\frac{1}{2}} \tilde{A} \alpha =
\tilde{S}_m^{-\frac{1}{2}} \tilde{A} E^T x' = \tilde{A}_R (x - \bar{x}),}
where we have defined the ``rotated'' EOF-space AK,
\eql{AR-EOF}{\tilde{A}_R = \tilde{S}_m^{-\frac{1}{2}} \tilde{A} E^T,}
for convenience.
Now the observation increments \eql{obs-inc-EOF}{\hat{y}_A - y_A =
\tilde{S}_m^{-\frac{1}{2}} \epsilon_{\alpha},} are unbiased, and have
covariance $I$.

Note that we include the EOF projection operator in our definition of
$\tilde{A}_R$ in \eqr{AR-EOF}.
This allows for the possibility of vertical localization and makes the
implementation in EOF or physical space similar.
Alternatively, one could determine $\alpha$ by projection (as in
\eqr{project-xhat} and \eqr{project-xa}), store rows of
$\tilde{S}_m^{-\frac{1}{2}} \tilde{A}$ with the observation $\hat{y}_A$
and then define the observation operator by first form in
\eqr{yA-EOF}.
This approach saves storage since it stores weights for EOF
coefficients, not temperatures, but is not amenable to localization.
Of course vertical localization may not be possible depending on the
structure of $\tilde{A}_R$.

\xx {Using the EOF representation for vertical interpolation\secl{EOF-vinterp}}

Vertical interpolation inevitably introduces some errors.
If we can somehow project the meteorological model vertical structure
directly onto the EOFs we could eliminate this source of error.
In this case we would use the first form of \eqr{yA-EOF},
$\tilde{S}_m^{-\frac{1}{2}} \tilde{A} \alpha$.
As described above we must interpolate the meteorological model values
to the levels used in the retrieval and then project using
\eqr{projection}.
However there are some complications that make this unworkable when
the grid definitions are very different, \eg, altitude relative to the
geoid vs.\ sigma. What follows is an alternative approach (that, as a
by-product, provides a method of vertical interpolation).

In general we want to convert a vertical profile of model temperatures
into the EOF coefficients defined relative to the retrieval vertical
coordinate.
For example, in the model, temperature might be stored as
$\sigma$-layer values, while the retrieval might operate on a fixed
$p$-level grid.
In this example, at some locations some of the $p$-levels might be
below model topography.
We require (and this is expected to always be the case) that there are
more model temperatures in the profile than there are EOF
coefficients.
To convert from model temperatures, $T$, to EOF coefficients, $\alpha$,
we determine $\alpha$ to fit the $T$ in a least squares sense.
That is we minimize
\eql{J}{J=\sum_i(T^*_i - T_i)^2,}
where $T_i$ are the background model temperatures and $T^*_i$ are
the model temperatures reconstructed from the $\alpha$.
Here we restrict the $T_i$ to just those model temperatures that can
be reconstructed from a specification of the EOF coefficients.
For example, any model temperatures above the uppermost temperature in
the retrieval vertical coordinate would not be included in the sum in
\eqr{J}.
The reconstructed temperatures are determined from $\alpha$ in two
steps.
(We don't actually perform these steps, but we must define them to
solve the least squares problem.)
First we reconstruct the temperature profile $x$ in the retrieval
vertical coordinate,
\eql{reconstruct}{x=E\alpha + \bar{x}.}
Second we interpolate to the model vertical coordinate using an
interpolation operator $V$ that is linear in the $x$,
\eql{vinterp}{T^* = V x = V (E\alpha + \bar{x}).}
Usually we will interpolate the temperatures linearly in pressure.
Note that in the example described above, $V$ will depend on the
model surface pressure and vary from location to location.
Instead of doing an actual interpolation, we must be able to determine
the matrix $V$.
To minimize $J$ \wrt\ the $\alpha$ we set the derivatives equal to
zero and find that\ssfootnote{Using Einstein notation (summation implied for
repeated indices), the reconstructed temperature at level $i$ is
written as $T^*_i = V_{ik}(E_{kn}\alpha_n + \bar{x}_k)$.
The derivative of this \wrt\ $\alpha_j$ may be written as
$E^T_{jk}V^T_{ki}$.
Then setting the derivative of $J$ to zero gives $0 = 2
E^T_{jk}V^T_{ki} (T^*_i - T_i)$.
Substituting for $T^*$ here and rearranging terms then gives
\eqr{ls-alpha}.}
\eql{ls-alpha}{E^T V^T V E \alpha = E^T V^T ( T - V \bar{x}).}
This is a linear equation of the form $M\alpha=b$ where the size of
$M$ might be $6\times6$, but $M$ will vary from location to location.
Note in \eqr{ls-alpha} that it is only the (presumably smooth)
$\bar{x}$ that is vertically interpolated.

Given the $\alpha$ determined this way we can then calculate $x$
according to \eqr{reconstruct}.\ssfootnote{Note that this produces
temperature values for all radiative transfer model grid points, even
below the surface in some cases.
Any values below the surface will not be used by the radiative
transfer model, but by construction these will be statistically
consistent with the rest of the profile.}
This would fit into either the AK method or the EOF-AK method.

\xx {Implementation details}

\xxx {Underground when using pressure as the vertical coordinate}

If the radiative transfer model (RTM) for $F$ uses a grid fixed in
pressure (or in altitude) then the lengths of the vectors and the
shapes of the matrices will vary from retrieval to retrieval as
topography and surface pressure vary from location to location.
We may simply truncate  those parts of the vectors and matrices that
are underground.
Operationally, we may define underground as those levels whose
corresponding rows of $K\equiv0$, \ie, levels that have no effect on
the calculated radiances.
Alternatively, we can keep all the vectors and matrices full size but
take care that underground levels are handled properly.
For example, underground rows of $K$ must be zero or must be strictly
ignored.
In particular, underground elements of $S_m$, $S_m^{-\frac{1}{2}}$,
and $S_a^{-1}$ should be reset to zero after they are calculated to
avoid the possible effects of accumulating round-off errors.

When using EOFs, underground levels in $A$, $G_y$ or $S_m$, and each
$x'$ should be zero.
Equivalently the variables in \eqr{AK-alpha} will be assured to be correct by
setting the rows of $E$ below the surface to zero.

\xxx {Generalized matrix inversion}

Since $S_a$ may be poorly conditioned it will generally be necessary
to use a truncated eigenvector decomposition to find the
(Moore-Penrose) inverse of $S_a$.
This is not necessary if $\hat{S}$ is provided along with $\hat{x}$.
Alternatively, if we are given $S_{\epsilon}$, $S_a$, and $K$, it is
possible to sidestep $\hat{S}$ entirely by using \eqr{Gy} and \eqr{Sq}.
Similarly to compute $S_m^{-\frac{1}{2}}$ from $S_m$ it may be
necessary to use a truncated eigenvector decomposition.

The idea here is that after we rotate to the eigenvector coordinate
system there are some coordinates that are so flat (have such small
eigenvalues) that they can be ignored. So we determine an inverse or
square root inverse that is formally correct in the full but
transformed space and then ignore (truncate) the near-singular dimensions.
The inverse matrix in the truncated transformed coordinate system is
in fact a proper Moore-Penrose pseudoinverse matrix in the original
coordinate system.
The correct choice of truncation would be the truncation used in the
retrieval scheme.
If an EOF truncation is not used in the retrieval scheme then the
correct choice of truncation may be apparent from the eigenvalue
spectrum.
Otherwise it will probably be sufficient to choose the truncation such
that at least 99.9\% of the variance is retained, \ie, such that the
sum of the retained eigenvalues is $\ge$ 99.9\% of the sum of all the
eigenvalues.

\emph{Eigen-decomposition.} Any positive definite matrix $S$, such as
a covariance matrix, may be written as
\eql{eigen-decomposition}{S=VGV^T,} where $V$ is a matrix of the
eigenvectors and $G$ is a diagonal matrix of the eignenvalues of $S$.

\emph{Inverse.} The pseudoinverse of $S$ is given by
\eql{inverse}{S^+=V G^{-1} V^T}
since
\eql{inverse-proof}{SS^+=V G V^T V G^{-1} V^T=V V^T=I.}
Note that $ V^T V=I$ regardless of the truncation, but $V V^T=I$ is
true only for the case of no truncation.
When we truncate both $V$ and $G$ we can easily demonstrate that $S^+$
satisfies the four required Moore-Penrose properties---$SS^+S=S, S^+SS^+=S^+,
(SS^+)^T= SS^+,\mbox{ and }(S^+S)^T= S^+S$.\ssfootnote{From \eqr{inverse-proof}
we have that $SS^+=VV^T$ which is symmetric, and reversing the roles of
$S$ and $S^+$ in \eqr{inverse-proof} gives the same result. Thus
$SS^+$ and $S^+S$ are symmetric and equal to $VV^T$. Clearly pre- or
post-multiplying $S$ or $S^+$ by $VV^T$ leaves them unchanged,
demonstrating the first and second properties. And because of
symmetry the third and fourth properties hold.}

\emph{Square root.} The square root of $S$ is given by
\eql{sqrt}{S^{\frac{1}{2}}=V G^{\frac{1}{2}} V^T}
since
\eql{sqrt-proof}{S^{\frac{1}{2}} S^{\frac{1}{2}}=V G^{\frac{1}{2}} V^T V G^{\frac{1}{2}} V^T = V G V^T
= S}
for any truncation.
According to \eqr{inverse} the pseudoinverse of $S^{\frac{1}{2}}$ is then
\eql{inverse-sqrt}{S^{\frac{1}{2}+}=V G^{-\frac{1}{2}} V^T.}

\xxx {Interface with retrieval and order of computation}

Consider how data flows through our procedures for a moment.
Given the retrieval, $\hat{x}$, once we know $A$ and $S_m$, we then
calculate $\hat{y}_A$, and $A_R$ from \eqr{yhatA}, and \eqr{AR}.
Usually, the retrieval will not provide $A$ and $S_m$.
Three possible pathways for the calculation of $A$ and $S_m$ are
described now.
(Refer to \tabr{pathways}.)
Inputs to the retrieval include $x_a$, $S_a$, and $S_{\epsilon}$.
In some situations, these three inputs are all constant.
The three sets of  outputs from the retrieval are $\hat{x}$ and (i)
$K$, or (ii) $\hat{S}$, or (iii) $\hat{S}$ and $A$.
In the first pathway, from the retrieval inputs $S_a$ and $S_{\epsilon}$ and the
specified retrieval output $K$ we can calculate in turn $S_q$, $G_y$, $A$, and
$S_m$ from \eqr{Sq}, \eqr{Gy}, the first form of \eqr{AK}, and
\eqr{Sm}.
This pathway requires only the inversion of $S_q$,
which should be well conditioned as mentioned before.
In the second pathway, using the retrieval input $S_a$, and the retrieval
output $\hat{S}$ we can calculate $A$
from the third form of \eqr{AK} and then $S_m$ from \eqr{SmA}.
In this case we need to invert $S_a$.
As pointed out by \citet{DeeEF+03} in cases where $x_a$ and $S_a$ are
fixed, this second pathway greatly reduces the data that must be
provided by the retrieval and $S_a^{-1}$ needs to be calculated only
once.
In the third pathway using the retrieval outputs $\hat{S}$ and $A$ we
can immediately calculate $S_m$ from \eqr{SmA}.
Note that in the second and third pathways the retrieval outputs are
all sized by the number of  retrieved quantities, and not the number
of radiances.

\renewcommand{\eqr}[1]{[Eq.~\ref{eq:#1}]}

\begin{table}
\begin{center}
\begin{tabular}{r|ccc} \hline
{\bfseries Pathway} & 1 & 2 & 3 \\ \hline \hline
\bfseries Interface & & & \\
Retrieval inputs &  $x_a, S_a, S_{\epsilon}$ & $x_a, S_a$ & $x_a$ \\
Retrieval outputs & $\hat{x}, K$ & $\hat{x}, \hat{S}$  & $\hat{x}, \hat{S}, A$ \\
\hline
{\bfseries Calculations} & & & \\
$S_q=$ & $K S_a K^T + S_{\epsilon}$~\eqr{Sq} & --- & --- \\
$G_y=$ & $S_a K^T S_q^{-1}$~\eqr{Gy} & --- & --- \\
$A=$ & $G_y K$~\eqr{AK} & $I-\hat{S} S_a^{-1}$~\eqr{AK} & --- \\
$S_m=$ & $G_y S_{\epsilon} G_y^T$~\eqr{Sm} & $A \hat{S}$~\eqr{SmA} & $A \hat{S}$~\eqr{SmA} \\ \hline
\end{tabular}
\end{center}
\tcaption{pathways}{Pathways to calculate $A$ and $S_m$. See text for discussion.}
\end{table}

\renewcommand{\eqr}[1]{Eq.~(\ref{eq:#1})}

When using EOFs, the above discussion still holds, but now we also
need $\bar{x}$ and $E$ to define the EOF representation.
Then once we have determined $A$ and $S_m$ as before, we can then calculate in
turn $\tilde{A}$, $\hat{\alpha}$, $\alpha_a$, $\tilde{S}_m$,
$\hat{y}_A$, and $\tilde{A}_R$ from \eqr{tildeA}, \eqr{project-xhat},
\eqr{project-xa}, \eqr{StildeM}, \eqr{yhatA-EOF}, and \eqr{AR-EOF}.

\xx {Algorithm implementation}

Here I sketch out \emph{one} implementation for interactive retrievals using
the AK-EOF method and the EOF vertical interpolation operator.
\begin{enumerate}

\item Assemble known constants: instrument description (channels,
radiance errors ($S_{\epsilon}$\ssfootnote{This should be
instrument noise plus forward problem errors
\citep[Eq.\ 3.11]{Rod00}. However in practice just
instrument noise is often used.}), geometry), EOF information (EOFs ($E$), climate mean
($\bar{x}$), truncation, eigenvalues), \ldots.

\item Build observation file with times, locations, radiances ($y$),
geometry, \ldots.

\item Use the background ensemble to calculate the prior mean and covariance,
$x_a$ and $S_a$, and add these to the observation file.
Note that $x_a$ and $\bar{x}$ are different in this case.\ssfootnote{We retain the
global EOFs in our process instead of trying to determine EOFs for
each retrieval---we want the EOFs to span all possibilities.
The alternative, a set of EOFs from a forecast, might be missing a
mode of variability actually present in reality and such an
inconsistency could adversely affect the retrievals.
I believe this issue---the fact that the ensemble does not have enough
variability to efficiently fit the radiances---needs some sort of fix in
radiance assimilation.
Quality control might help, but is the wrong solution---getting rid of
good but unusual data because the model is inadequate.}

\item Perform retrieval. Add $\hat{x}$ and $\hat{S}$ in the retrieval output
file.\ssfootnote{We would also save $K$ and $\hat{y}=F(\hat{x})$ as defined in
\secr{forwardErrors} to be used later in \eqr{yb} if we are applying
the \citet{DesBC+05} method.}

\item Post-process retrieval. Following pathway 2, calculate in turn $A$, $S_m$,
$\tilde{S}_m$, $\tilde{A}$, $\hat{\alpha}$, $\alpha_a$, $\hat{y}_A$
(from \eqr{yhatA-EOF}), and $\tilde{A}_R$.
Save $\hat{y}_A$ and $\tilde{A}_R$ in (LETKF) obs-data structure.
The $i$th element of $\hat{y}_A$ is one observation.
It is associated with the $i$th row of $\tilde{A}_R$.

\item Data selection.
For vertical data localization use the absolute value of the  $i$th row of
$A_R$ as weights in the same way that we now localize radiances.

\item Vertical interpolation.
After interpolating the model grids to the map location and time of
the observation, use the procedure of \secr{EOF-vinterp} to determine
$\alpha$ and then $x$ using \eqr{ls-alpha} and \eqr{reconstruct}.

\item Simulate observation.
Calculate $y_A$ according to \eqr{yA-EOF}.\ssfootnote{At this point we
could also calculate the background radiance, $y_b$, as described in
\secr{forwardErrors} and save it in the observation data structure to
be used later to define the analysis and background ``observations''
for use in the \citet{DesBC+05} method.}

\end{enumerate}
The case outlined above is the most complicated one.
Other implementations of the methods described here could follow this
plan with appropriate simplifications. 

\xx {Acknowledgments}

I thank Dan Gombos, Janusz Eluskiewicz, Steven Greybush, Matthew J. Hoffman,
Kayo Ide, Alan Lipton, Eugenia Kalnay, Jean-Luc Moncet, David Kuhl, Thomas
Nehrkorn, Vivienne Payne, and R. John Wilson for helpful discussions.
This work supported in part by NASA grant NNX07AN97G.


\end{document}